\documentclass[english,twocolumn]{revtex4-1}
\usepackage[T1]{fontenc}
\usepackage[latin9]{inputenc}
\usepackage{verbatim}
\usepackage{amsmath}
\usepackage{amssymb}
\usepackage{graphicx}
\usepackage{babel}
\begin{document}
\title{Relationship between the wave function of a magnet \\
and its static structure factor}
\author{Jorge Quintanilla}
\address{Physics and Astronomy, Division of Natural Sciences, University of
Kent, Canterbury, CT1 7NH, United Kingdom}
\date{19 July 2022}
\begin{abstract}
We state and prove two theorems about the ground state of magnetic
systems described by very general Heisenberg-type models and discuss
their implications for magnetic neutron scattering. The first theorem
states that two models cannot have the same correlator without sharing
the corresponding ground states. According to the second theorem,
an $N$-qubit wave function cannot reproduce the correlators of a
given system unless it represents a true ground state of that system.
We discuss the implications for neutron scattering inverse problems.
We argue that the first theorem provides a framework to understand
neutron-based Hamiltonian learning. Furthermore, we propose a variational
approach to quantum magnets based on the second theorem where a representation
of the wave function (held, for instance, in a neural network or in
the qubit register of a quantum processor) is optimised to fit experimental
neutron-scattering data directly, without the involvement of a model
Hamiltonian. 
\end{abstract}
\maketitle
The Rayleigh-Ritz variational principle states that the ground state
wave function of a quantum system is an absolute minimum of the energy.
It provides the theoretical underpinning of many successful approaches
to the quantum many-body problem including Density Functional Theory
(DFT) \citep{capelle2006bird}, Variational Monte Carlo methods \citep{Scherer2017},
the BCS theory of superconductors \citep{Rickayzen1969} and the Laughlin
theory of the fractional quantum Hall effect \citep{laughlin1983anomalous}
to name a few cases. More recently it has been used to find optimal
representations of wave functions using quantum computers \citep{Tilly2021,Rattew2020}
and neural networks \citep{ANN1}. Such theories start with a model
Hamiltonian $\hat{H}$ and proceed by minimizing the energy $\left\langle \Psi\left|\hat{H}\right|\Psi\right\rangle $
to obtain the wave function $\Psi$. The Rayleigh-Ritz variational
principle ensures that no wave function can yield a lower value of
the energy than the system's true ground state. Once the wave function
is known, it is is straight-forward to predict expectation values
of observables. Very often, however, $\hat{H}$ is not known \emph{a
priori}. In such instances $\hat{H}$ has to be found from experimental
data. That involves a laborious and ill-posed inverse problem: multiple
candidate Hamiltonians must be studied until one is found that predicts
the experimentally-determined value of a set of observables. In general
there is no guarantee of uniqueness of $\hat{H}$ or $\Psi$ for a
given data set. Here we consider the inverse problem for the magnetic
structure factor of a magnetic insulator (in particular, one described
by an anisotropic Heisenberg model, which covers a vast range of real
materials). We show that, for systems that have non-degenerate, distinct
ground states, there is a one-to-one correspondence between the structure
factors, the model Hamiltonian and the ground state wave function.
We then address the implications of degeneracy, Hamiltonians with
the same ground state, and excitations, and discuss the implications
for neutron scattering. 
\begin{figure}
\includegraphics[width=1\columnwidth]{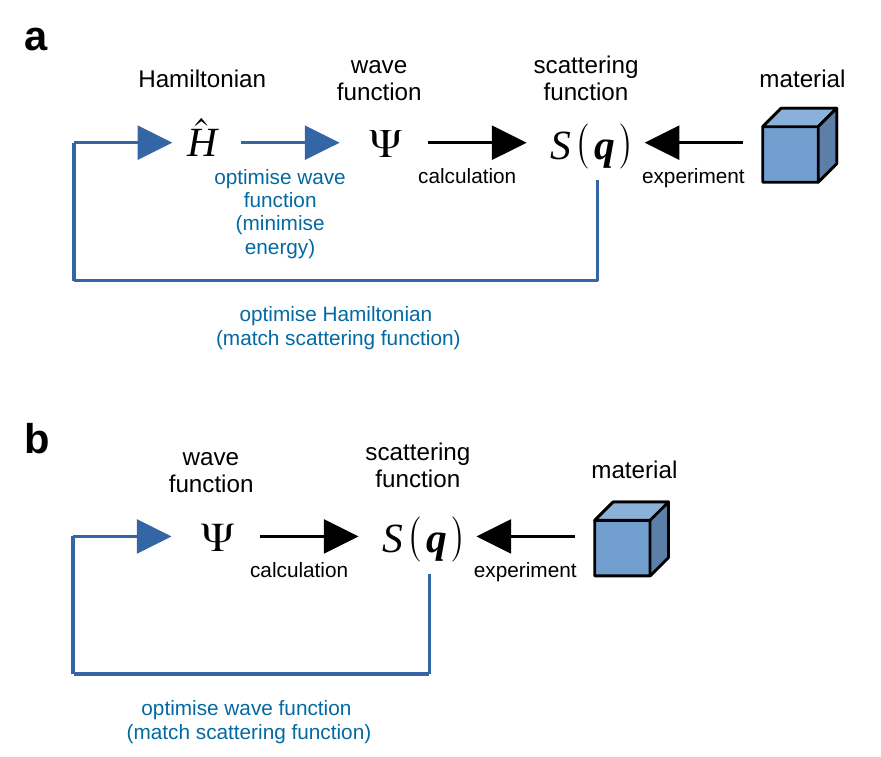}

\caption{\label{fig:-Two-versions}Two versions of the diffuse magnetic neutron
scattering inverse problem: given the scattering function $S\left(\mathbf{q}\right)$
of a real material, in Hamiltonian learning (a) the aim is to determine
a model Hamiltonian $\hat{H}$ whose wave function $\Psi$ will describe
$S\left(\mathbf{q}\right)$ satisfactorily. In quantum tomography
(b) one tries to determine the wave function $\Psi$ directly. }
\end{figure}

Our results have several direct implications for the study of magnetic
insulators using neutron scattering, specifically for the neutron
scattering inverse problems described schematically in Fig.~\ref{fig:-Two-versions}.
Firstly, as we argue below, Theorem 1 puts the Hamiltonian-learning
problem (Fig.~\ref{fig:-Two-versions}~a) on a firmer footing and
will help the design of efficient solutions, for instance ones exploiting
machine learning \citep{Samarakoon2020}. Secondly, Theorem 2 suggests,
and supports, new variational methods where the wave function is optimized
to describe the experimental data, obviating the need to minimize
the energy of a model Hamiltonian (Fig.~\ref{fig:-Two-versions}~b).
This provides an alternative to existing methods used to obtain the
ground state of a Heisenberg-type magnet, for instance those based
on neural-network \citep{ANN1} or quantum-processor \citep{Tilly2021}
representations. Many such methods are based on minimizing the energy
for a given model. The new methods will be appropriate when the model
is not yet known but experimental structure factor information is
available. In those circumstances, working with the wave function
directly has the advantage of involving a single optimization loop
rather than two nested ones (compare Fig.~\ref{fig:-Two-versions}a
to \ref{fig:-Two-versions}b). In analogy with the Rayleigh-Ritz variational
principle, our second theorem guarantees that no wave function other
than a true ground-state wave function of the system under investigation
can yield a better fit to the data. Finally, our results suggest that
every ground-state property of the system is contained in the structure
factors. This has important implications for efforts to quantify quantum
entanglement from experimental neutron scattering data \citep{Brukner2006,marty2014quantifying,Laurell2021}
and justifies the reduction of measures of entanglement to functions
of correlators \citep{Amico2004,Baroni2007}.

The work presented here has to be seen in the context of recently-developed
methods for the determination of model Hamiltonians from local measurements
\citep{Chertkov2018,Greiter2018,Bairey2019,Qi_2019,Anshu2021}. Interestingly,
a main thrust of such works, which are usually concerned with systems
where qubits need to be addressed individually, is the optimisation
of the scaling of the number and type of measurements required with
the size of the system and the range of interactions. In contrast,
our approach relies on the static magnetic structure factor $S_{\alpha,\beta}\left(\mathbf{q}\right)$
which contains information about all two-point correlators and can
be determined experimentally with the same effort irrespective of
system size or range of interactions. 

Our starting assumption is that the physical system under experimental
investigation can be described by an anisotropic Heisenberg model:
\begin{equation}
\hat{H}=\sum_{i,j}\sum_{\alpha,\beta}J_{i,j}^{\alpha,\beta}\hat{S}_{i}^{\alpha}\hat{S}_{j}^{\beta}.\label{eq:H}
\end{equation}
Here $i,j=1,2,\ldots,N$ represent atomic sites, whose positions $\mathbf{R}_{i},\mathbf{R}_{j}$
we assume to be known. $\hat{S}_{i}^{\alpha}$ represents the $\alpha^{\underline{\text{th}}}$
component of the spin operator for the magnetic moment at the $i^{\underline{\text{th}}}$
atomic site ($\alpha=x,y,z$; we assume each spin component is defined
with reference to some local axes defined on each site). We assume
the spin quantum number at each site is $S=1/2$ in what follows but
the results can be generalised to arbitrary $S$ straight-forwardly.
$J_{i,j}^{\alpha,\beta}$ is an exchange constant describing the interaction
between the $\alpha^{\underline{\text{th}}}$ component of the spin
at the $i^{\underline{\text{th}}}$ site of a given lattice and the
$\beta$ component of the spin in the $j^{\underline{\text{th}}}$
site. The terms with $i=j$ describe site anisotropy (e.g. easy planes
or easy axes). The dependence of $J_{i,j}^{\alpha,\beta}$ on $i,j,\alpha,$
and $\beta$ is entirely arbitrary. The model of Eq.~(\ref{eq:H})
can thus describe a very broad range of magnetic models in arbitrary
dimensions with and without translational invariance, including among
others the Ising model \citep{Ising1925}, XY model \citep{wang2001entanglement},
and Kitaev model to name but a few \citep{kitaev2006anyons}. Models
of this type are believed to describe well the physics of many materials
from single-molecule magnets \citep{Christou2000} through infinite-chain
compounds \citep{Laurell2021} to three-dimensional quantum spin ices
\citep{gingras2014quantum} and other spin liquids \citep{Broholm2020}.
The observable quantity of interest is the two-point magnetic correlator
\begin{equation}
\rho_{i,j}^{\alpha,\beta}\left[\Psi\right]\equiv\left\langle \Psi\left|\hat{S}_{i}^{\alpha}\hat{S}_{j}^{\beta}\right|\Psi\right\rangle .\label{eq:rho}
\end{equation}
The correlator is obviously a single-valued functional of the wave
function $\Psi$. As shown in Appendix \ref{sec:Equivalence-between-correlators}
this quantity is readily obtainable in condensed matter systems\emph{
}through neutron scattering mesurements of the static structure factor
$S_{\alpha,\beta}\left(\mathbf{q}\right)$. 

\begin{figure}
\includegraphics[width=1\columnwidth]{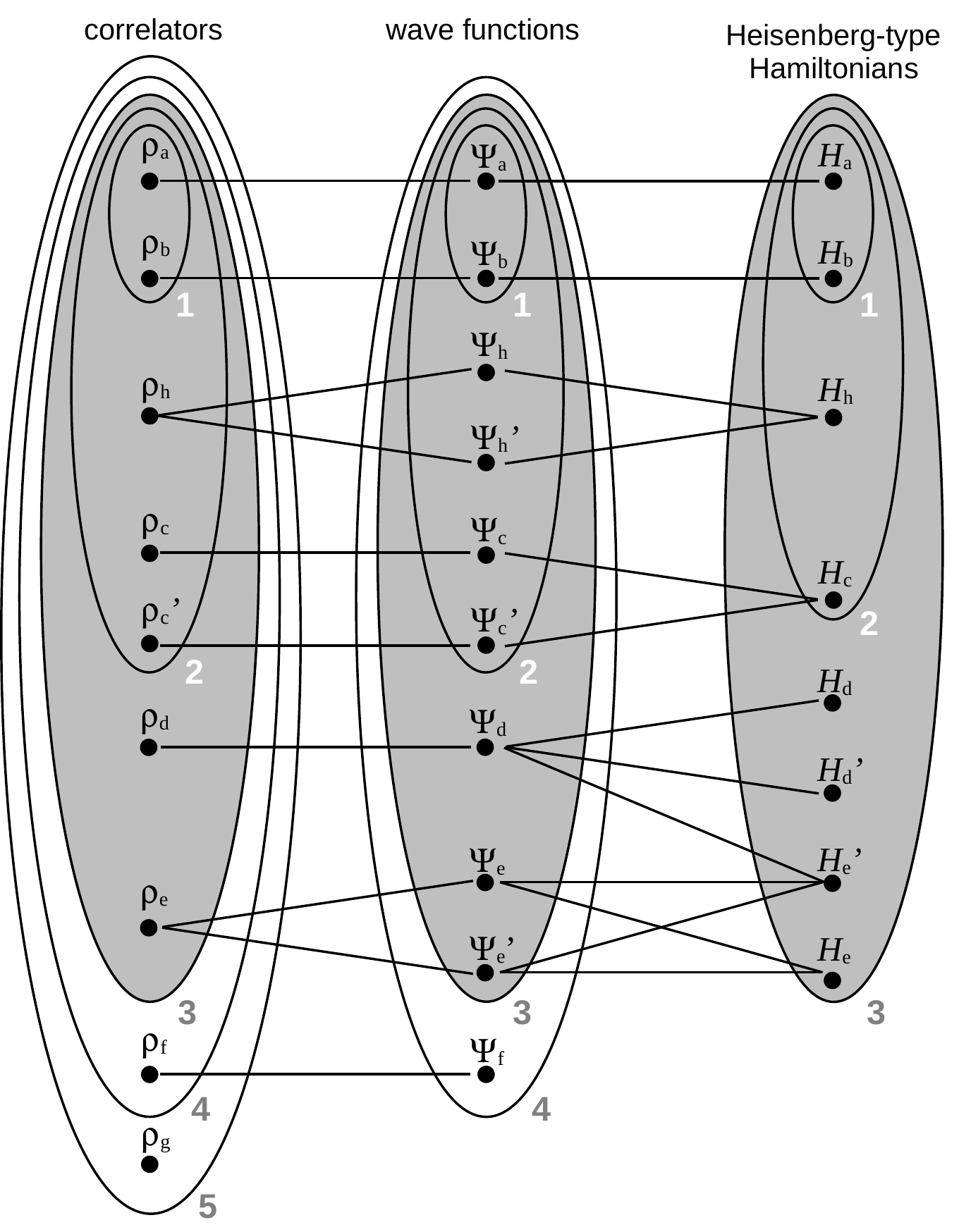}%

\caption{\label{fig:Possible-relationships-between}Schematic illustration
of possible relationships between 2-point correlation functions $\rho$,
$N$-qubit wave functions $\Psi$, and general spin-1/2 anisotropic
Heisenberg Hamiltonians $H$ (see main text). }
\end{figure}
The situation we have in mind is one where the ground-state correlator
$\rho_{i,j}^{\alpha,\beta}\left[\Psi_{0}\right]$ has been obtained
experimentally but neither the Hamiltonian $\hat{H}$ nor the wave
function $\Psi_{0}$ are known. We wish to prove two closely related
theorems that impose constraints on $\hat{H}$ and $\Psi_{0}$:
\begin{description}
\item [{Theorem~1}] Two Hamiltonians $\hat{H},\hat{H}'$ cannot have the
same ground-state correlator without sharing the corresponding ground
states. 
\item [{Theorem~2}] Any $N$-qubit wave function $\Psi$ that can reproduce
the ground-state correlator of $\hat{H}$ represents a ground state
of $\hat{H}$.
\end{description}
The implications of these two theorems for the relationship between
correlators, wave functions, and Hamiltonians are illustrated in Fig.~\ref{fig:Possible-relationships-between}.
For non-degenerate Hamiltonians, Theorem 1 is a particular case of
a more general theorem proven by Ng~\citep{Ng1991}. Here we extend
it to the important case of Hamiltonians with degenerate ground states
and discuss the implications of Hamiltonians sharing a ground state
for Hamiltonian learning using neutron-scattering data (Fig.~\ref{fig:-Two-versions}a).
The latter discussion will be supported by a generalisation of Theorem
1 to excited states (Appendix~\ref{sec:Can-two-different}). Theorem
2, on the other hand, is a consequence of the fact that the ground
state maximises the reconstruction entropy given a set of local measurements
\citep{Swingle2014}. The theorems are also closely related to the
convexity of the set of all 2-point correlators of $N$-qubit states
\citep{Schwerdtfeger2009,Zauner2016}. Note that our second theorem
does not require all correlators to be representable by $N$-qubit
wave functions (exemplified by $\rho_{\text{g}}$ Fig.~\ref{fig:Possible-relationships-between})
nor is it restricted to trial wave functions that are ground states
of Heisenberg-type Hamiltonians ($\rho_{\text{f}}$ in the Figure).
We will discuss the implications of this second theorem for neutron-based
quantum tomography (Fig.~\ref{fig:-Two-versions}b). Finally we note
that Theorem~1 can be deduced from Theorem 2 however for clarity
we will prove both theorems independently.

Our approach to proving Theorems 1 and 2 is inspired by the DFT formalism
for Heisenberg models developed by L{\'i}bero and Capelle \citep{libero2003spin}.
Our aims, however, are quite different. The latter work (like other
DFT formalisms for lattice models \citep{libero2003spin,Wu2006,Coe2015})
is an energy-minimization variational theory closely modeled on the
original DFT for electrons in solids \citep{capelle2006bird}. In
Density Functional Theories generally, the aim is to show that the
energy is a functional of a density-like quantity (in the case of
Ref.~\citep{libero2003spin}, the local magnetisation). One then
splits the energy into two parts, one that is ``universal'' and
another that depends on local fields. In order for this to be useful,
it is necessary to have exact results for the universal function and
motivated approximations for the field-dependent contribution. Here
our primary quantity is not a density but a correlator, and we are
not interested in splitting the energy into one contribution that
is known and another that is to be approximated. Instead we treat
the energy as a single unit and are interested in proving that only
one ``universal'' Hamiltonian is compatible with a given set of
correlators. In practice, applications of our approach involve the
optimization of the match to experimental data, rather than the minimization
of the energy. Moreover, we will work in the absence of a known model
Hamiltonian, rather than using knowledge of one model (e.g. a translationally-invariant
Heisenberg model) to approximately solve another (e.g. the same model
but with an impurity potential). 

\emph{Proof of Theorem 1.- }Inspired by the original proof of the
Hohenberg-Kohn theorem of DFT, we will proceed by \emph{reductio ad
absurdum}. Suppose there are two distinct Hamiltonians $\hat{H}$
and $\hat{H}'$, with different exchange interaction functions $J_{i,j}^{\alpha,\beta}$
and ${J'}{}_{i,j}^{\alpha,\beta}$ and different ground states $|\Psi_{0}\rangle,|\Psi_{0}'\rangle$,
respectively, that give the same correlator $\rho_{i,j}^{\alpha,\beta}$:
\begin{equation}
\rho_{i,j}^{\alpha,\beta}\left[\Psi_{0}\right]=\rho_{i,j}^{\alpha,\beta}\left[\Psi_{0}'\right]\text{ for all }i,j,\alpha,\beta.\label{eq:rho_the_same}
\end{equation}
We will first consider the case when the two ground states are non-degenerate.
In this case the ground-state energy obtained from the first Hamiltonian
is 
\begin{align}
E_{0}=\left\langle \Psi_{0}\left|\hat{H}\right|\Psi_{0}\right\rangle  & <\left\langle \Psi'_{0}\left|\hat{H}\right|\Psi'_{0}\right\rangle \label{eq:ineq1}\\
 & =\left\langle \Psi'_{0}\left|\hat{H}-\hat{H}'\right|\Psi'_{0}\right\rangle +\left\langle \Psi'_{0}\left|\hat{H}'\right|\Psi'_{0}\right\rangle \\
 & =\sum_{i,j}\sum_{\alpha,\beta}\left(J_{i,j}^{\alpha,\beta}-{J'}{}_{i,j}^{\alpha,\beta}\right)\rho_{i,j}^{\alpha,\beta}\left[\Psi_{0}'\right]+E_{0}'
\end{align}
where the inequality is due to the Rayleigh-Ritz variational principle.
Similarly the ground state energy obtained from the second Hamiltonian
is 
\begin{align}
E_{0}'=\left\langle \Psi_{0}'\left|\hat{H}'\right|\Psi_{0}'\right\rangle  & <\left\langle \Psi_{0}\left|\hat{H}'\right|\Psi_{0}\right\rangle \label{eq:ineq2}\\
 & =\left\langle \Psi_{0}\left|\hat{H}'-\hat{H}\right|\Psi_{0}\right\rangle +\left\langle \Psi_{0}\left|\hat{H}\right|\Psi_{0}\right\rangle \\
 & =\sum_{i,j}\sum_{\alpha,\beta}\left({J'}{}_{i,j}^{\alpha,\beta}-J_{i,j}^{\alpha,\beta}\right)\rho_{i,j}^{\alpha,\beta}\left[\Psi_{0}\right]+E_{0},
\end{align}
Adding the two inequalities we obtain 
\begin{align*}
E_{0}'+E_{0}< & \sum_{i,j}\sum_{\alpha,\beta}\left({J'}{}_{i,j}^{\alpha,\beta}-J_{i,j}^{\alpha,\beta}\right)\\
 & \times\left\{ \rho_{i,j}^{\alpha,\beta}\left[\Psi_{0}\right]-\rho_{i,j}^{\alpha,\beta}\left[\Psi_{0}'\right]\right\} +E_{0}+E_{0}'.
\end{align*}
Using now our assumption (\ref{eq:rho_the_same}) this reduces to
\begin{equation}
E_{0}'+E_{0}<E_{0}+E_{0}'\label{eq:contrad1}
\end{equation}
which is absurd. Thus our initial assumption must be incorrect: two
Heisenberg-type Hamiltonians with different exchange interaction constants
and distinct, non-degenerate ground states can never give the same
correlator. In other words, for non-degenerate Hamiltonians that do
not share ground states the exchange interaction function is a single-valued
functional $J_{i,j}^{\alpha,\beta}\left[\rho\right]$ of the correlator
$\rho_{i,j}^{\alpha,\beta}$. This is illustrated by the one-to-one
correspondence between $\left\{ H_{\text{a}},H_{\text{b}}\right\} \text{ and }\left\{ \rho_{\text{a}},\rho_{\text{b}}\right\} $
in Fig.~\ref{fig:Possible-relationships-between}. 

We note that our proof relies on the assumption that $|\Psi_{0}\rangle\neq|{\Psi'}_{0}\rangle$
since otherwise the strict inequalities (\ref{eq:ineq1},\ref{eq:ineq2})
become equalities. In other words, if $\hat{H}$ and $\hat{H}'$ share
their unique ground state the theorem does not apply. This is illustrated
by the one-to-many correspondence between $\rho_{\text{d}}$ and $\left\{ H_{\text{d}},H_{\text{d}}',H_{\text{e}}'\right\} $
in Fig.~\ref{fig:Possible-relationships-between}. Although this
might appear to be a serious limitation  for Hamiltonian learning
using neutron scattering it may not be as important in practice, as
we discuss below. 

In the above paragraphs we explicitly assumed that the ground states
of $\hat{H}$ and $\hat{H}'$ are non-degenerate. In order to prove
Theorem 1 we need to relax that assumption. Let us first consider
the case when the ground state of one of the Hamiltonians (which we
take to be $\hat{H}$ without loss of generality) is degenerate, while
that of the other Hamiltonian remains non-degenerate. Then the first
of the above two inequalities (\ref{eq:ineq1},\ref{eq:ineq2}) is
not strict, as there is always the possibility that $|\Psi_{0}'\rangle$
happens to be a ground state of $\hat{H}$ as well as being the unique
ground state of $\hat{H}'.$ In that case, Theorem 1 would be violated,
because $|\Psi_{0}\rangle$ would not be a ground state of $\hat{H}'$
yet it would have the same correlators as $|\Psi_{0}'\rangle,$ which
is. Barring that possibility, the arguments above hold, so we only
need to consider that special case. In the special case we have 
\begin{align}
E_{0} & =\sum_{i,j}\sum_{\alpha,\beta}\left(J_{i,j}^{\alpha,\beta}-{J'}{}_{i,j}^{\alpha,\beta}\right)\rho_{i,j}^{\alpha,\beta}\left[\Psi_{0}'\right]+E_{0}'\label{eq:spec1}
\end{align}
and
\begin{equation}
E_{0}'<\sum_{i,j}\sum_{\alpha,\beta}\left({J'}{}_{i,j}^{\alpha,\beta}-J_{i,j}^{\alpha,\beta}\right)\rho_{i,j}^{\alpha,\beta}\left[\Psi_{0}\right]+E_{0}\label{eq:spec2}
\end{equation}
When we add the equality (\ref{eq:spec1}) to this inequality (\ref{eq:spec2})
we still arrive at the same contradiction as before, (\ref{eq:contrad1}).
This proves that Theorem 1 also applies when one of the two Hamiltonians
is degenerate.

Let us now consider the case when \emph{both} Hamiltonians have ground-state
degeneracy. Then there is a new possibility, namely that $|\Psi_{0}\rangle$
is a ground state of $\hat{H}'$ \emph{and} $|\Psi'_{0}\rangle$ is
a ground state of $\hat{H}$ (all other possibilities have already
been covered above). In that case all the above inequalities become
equalities and we do not arrive at a contradiction. However, in this
case all the ground states of $\hat{H}$ and $\hat{H}'$ leading to
the same correlators are shared so the premise of the theorem is not
satisfied. In summary, degenerate Hamiltonians with the same correlators
must share \emph{all }the corresponding ground states, as a single
ground state of one which is not also a ground state of the other
suffices to generate a contradiction. This is illustrated by $\rho_{\text{e}}$
in Fig.~\ref{fig:Possible-relationships-between}: the same correlator
can be generated by the ground state of $H_{\text{e}}$ and by that
of $H_{\text{e}}'$ but then the ground state of $H_{\text{e}}'$
must also be a ground state of $H_{\text{e}}$, and \emph{vice versa}
(it is possible, on the other hand, for ground states leading to \emph{different}
correlators not to be shared, as illustrated by the correlator $\rho_{\text{d}}$
of $H_{\text{e}}'$). This concludes our proof of Theorem 1.$\square$

Before proving Theorem 2 we shall discuss the implications of Theorem
1 for Hamiltonian learning using neutron scattering. First of all,
we address the question of shared ground states. %
{} An example of this would be two spin-1/2 ferromagnetic Ising models
differing only by an overall multiplicative factor. In both cases,
the ground state is a classical state where all the spins are pointing
along the positive or negative direction of the quantization axis.
The correlators are therefore identical in the ground state. It is
therefore impossible to discriminate between these two models from
a measurement of the ground-state correlators. It would be tempting
to venture that this limitation of Theorem 1 can be trivially circumvented
by normalising the parameters of trial Hamiltonians. However, this
would only take care of certain instances of shared ground states,
known \emph{a priori}. We cannot discard non-trivial cases. On the
other hand, it is straight-forward to generalise our proof of Theorem
1 to show that it holds for any excited state $|\Psi_{n}\rangle$
as well as for the ground state (see Appendix \ref{sec:Can-two-different}).
Therefore, two Hamiltonians that are physically distinct but share
a ground state could be told apart by probing their low-energy excitations.
Indeed any real condensed-matter experiment will take place at finite
temperature with the measured correlator corresponding to a thermal
superposition $\sum_{n}Z^{-1}\exp\left(-E_{n}/k_{B}T\right)\rho_{i,j}^{\alpha,\beta}\left[\Psi_{n}\right]$.
For two Hamiltonians $\hat{H}$ and $\hat{H}'$ to give the same result
at any arbitrary temperature one would therefore require all eigenstates
$|\Psi_{n}\rangle$ and all the eigenvalues $E_{n}$ to coincide.
In that case, the two Hamiltonians are identical: $\hat{H}=\hat{H}'=\sum_{n}E_{n}|\Psi_{n}\rangle\langle\Psi_{n}|$.
Thus this limitation of Theorem 1 may not, in practice, limit its
applicability to Hamiltonian learning in neutron scattering experiments.\footnote{We also note that this limitation of Theorem 1 does not affect Theorem
2, below.}\footnote{Murta and Fern{\'a}ndez-Rossier have recently generalised Theorem
1 to finite temperatures~\citep{Murta2022}. }

The above discussion suggests that the inverse problem of deducing
the Hamiltonian from the correlators may indeed be well-defined, as
long as we know that the material under investigation is described
by a model of the form in Eq.~(\ref{eq:H}). That could provide a
natural explanation for the success of a recent machine learning based
approach to this problem \citep{Samarakoon2020}. In that reference
an auto-encoder was trained using simulations of the neutron scattering
function $S_{\alpha,\beta}\left(\mathbf{q}\right)$ obtained for a
family of candidate Hamiltonians {[}for completeness, we have offered
a proof of the equivalence between knowledge of $S_{\alpha,\beta}\left(\mathbf{q}\right)$
and of $\rho_{i,j}^{\alpha,\beta}$ in Appendix \ref{sec:Equivalence-between-correlators}{]}.
The auto-encoder thus trained can be used to generate a low-dimensional
latent space on which experimental data can be projected, effectively
finding an optimal model Hamiltonian. Though in principle that inverse
problem is ``ill-defined'' \citep{Samarakoon2020}, our formal results
about the ground and excited states of Heisenberg-type Hamiltonians
strongly suggest that there may only be one solution. We note that
the work in Ref.~\citep{Samarakoon2020} dealt with classical models
however similar dimensionality-reduction has been shown for quantum
models using closely-related Principal Component Analysis \citep{Twyman2021}.

Several recent works have discussed the determination of model Hamiltonians
using local measurements \citep{Chertkov2018,Greiter2018,Bairey2019,Qi_2019,Anshu2021}.
While such methods are well suited to artificial systems such as quantum
simulators they are not readily applicable to experimental data on
condensed matter systems. Specifically in most cases \citep{Chertkov2018,Greiter2018,Bairey2019,Qi_2019}
they require the covariance matrix which in turn relies on 4-point
correlators of the form $\rho_{i,j,i',j'}^{\alpha,\beta,\alpha',\beta'}\left[\Psi_{0}\right]\equiv\left\langle \Psi_{0}\left|\hat{S}_{i}^{\alpha}\hat{S}_{j}^{\beta}\hat{S}_{i'}^{\alpha'}\hat{S}_{j'}^{\beta'}\right|\Psi_{0}\right\rangle $
where $i$ and $i'$ are sites that are not linked by a direct interaction
and $j$ and $j'$ are sites that interact with $i$ and $i'$, respectively.
Such higher-order correlators are not readily accessible through neutron
scattering. In contrast, for periodic systems all the 2-point correlators
$\rho_{i,j}^{\alpha,\beta}\left[\Psi_{0}\right]$ can be determined
in a neutron scattering experiment (see Appendix \ref{sec:Equivalence-between-correlators}).
As an added benefit, such approach yields all the required information
using a ``single shot'' global measurement irrespective of the size
of the system or the range of spin-spin interactions. Thus in the
case of condensed matter systems it is not necessary to devise more
sophisticated observables in order to improve sampling efficiency,
as was proposed recently \citep{Anshu2021}. Effectively, neutron
scattering integrates a large number of local measurements into a
single function $S\left(\mathbf{q}\right)$ that needs to be fitted
(see Appendix \ref{sec:Equivalence-between-correlators}) - a sort
of analogue parallel computation.

\emph{Proof of Theorem 2.-} In the preceding paragraphs we proved
that, for any set of Heisenberg-type Hamiltonians that do not share
ground states, the exchange constants $J_{i,j}^{\alpha,\beta}$ are
single-valued functionals of the correlators $\rho_{i,j}^{\alpha,\beta}$
(sets labelled ``2'' in Fig.~\ref{fig:Possible-relationships-between}).
With the additional constraint that the Hamiltonians have non-degenerate
ground states (sets labelled ``1'') the ground state $|\Psi_{0}\rangle$
is in turn fixed by the choice of $J_{i,j}^{\alpha,\beta}$. Thus,
in this case $|\Psi_{0}\rangle$ is also uniquely determined by $\rho_{i,j}^{\alpha,\beta}$.
More generally, Theorem~1 implies that the only ground states of
Heisenberg-type Hamiltonians (including Hamiltonians with degenerate
and/or shared ground states) that are compatible with the ground-state
correlator of a given model are also ground states of that same model.
The various possibilities are shown in Fig.~\ref{fig:Possible-relationships-between}:
each correlator in the shaded area (sets labelled ``3'' in the figure)
uniquely identifies a non-degenerate ground state ($\rho_{\text{a}}\to\Psi_{\text{a}},\rho_{\text{b}}\to\Psi_{\text{b}},\rho_{\text{c}}\to\Psi_{\text{c}},\rho_{\text{c}}'\to\Psi_{\text{c}}',\rho_{\text{d}}\to\Psi_{\text{d}}$)
or a set of degenerate ground states of either one Hamiltonian ($\rho_{\text{h}}\to\left\{ \Psi_{\text{h}},\Psi_{\text{h}}'\right\} $)
or of a set of Hamiltonians that share all those states ($\rho_{\text{e}}\to\left\{ \Psi_{\text{e}},\Psi_{\text{e}}'\right\} $)
\footnote{We note that as a direct consequence of Theorem 1 the relationship
between two ground states of Heisenberg-type models that exists by
virtue of them sharing the correlator defines a partition of that
set and can, therefore, be used to define equivalence classes for
ground states. The same is not true of the Hamiltonians themselves
as some degenerate Hamiltonians will have ground states with different
correlators - the latter point is illustrated by $H_{\text{d}}'$
in Fig.~\ref{fig:Possible-relationships-between}.}. This is essentially the same as Theorem 2 except for the constraint
that the trial wave functions must be ground states of Heisenberg-type
Hamiltonians. To prove Theorem~2 we need to show that the result
holds even without that constraint. In other words, we need to prove
that in Fig.~\ref{fig:Possible-relationships-between} there can
be no lines linking correlators in the shaded area (set labelled ``3''
on the left side of the figure) to wave functions in the unshaded
area (set labelled ``4'' in the middle). Again, we proceed by \emph{reductio
ad absurdum. }Let us assume that there is a state $|\tilde{\Psi}\rangle$
that gives the same correlator as $|\Psi_{0}\rangle$: 
\begin{equation}
\rho_{i,j}^{\alpha,\beta}\left[\tilde{\Psi}\right]=\rho_{i,j}^{\alpha,\beta}\left[\Psi_{0}\right]\text{ for all }i,j,\alpha,\beta.\label{eq:rho_the_same_tilde}
\end{equation}
Let us further assume that $|\tilde{\Psi}\rangle$ is \emph{not }the
ground state of $\hat{H}$. There are two possibilities: either it
is the ground state of some other Heisenberg-type Hamiltonian or it
is not the ground state of a Heisenberg Hamiltonian at all. Below
we will not assume either case, so our proof will cover both instances.
By the Rayleigh-Ritz variational principle, we know that $|\Psi_{0}\rangle$
gives the absolute minimum of the energy,  which implies  
\begin{equation}
E_{0}\equiv\left\langle \Psi_{0}\left|\hat{H}\right|\Psi_{0}\right\rangle \leq\left\langle \tilde{\Psi}\left|\hat{H}\right|\tilde{\Psi}\right\rangle .\label{eq:ene_expec_vals}
\end{equation}
Using Eqs.~(\ref{eq:H}) and (\ref{eq:rho}) we can write this as
\[
E_{0}\equiv\sum_{i,j}\sum_{\alpha,\beta}J_{i,j}^{\alpha,\beta}\rho_{i,j}^{\alpha,\beta}\left[\Psi_{0}\right]\leq\sum_{i,j}\sum_{\alpha,\beta}J_{i,j}^{\alpha,\beta}\rho_{i,j}^{\alpha,\beta}\left[\tilde{\Psi}\right].
\]
Let us now consider separately the two cases when the two expectation
values of $\hat{H}$ in Eq.~(\ref{eq:ene_expec_vals}) are different
and when they are equal. Let us first consider the case when they
are different: 
\[
E_{0}\equiv\left\langle \Psi_{0}\left|\hat{H}\right|\Psi_{0}\right\rangle <\left\langle \tilde{\Psi}\left|\hat{H}\right|\tilde{\Psi}\right\rangle .
\]
Then we have 
\begin{equation}
\sum_{i,j}\sum_{\alpha,\beta}J_{i,j}^{\alpha,\beta}\rho_{i,j}^{\alpha,\beta}\left[\Psi_{0}\right]<\sum_{i,j}\sum_{\alpha,\beta}J_{i,j}^{\alpha,\beta}\rho_{i,j}^{\alpha,\beta}\left[\tilde{\Psi}\right].\label{eq:contrad_tilde}
\end{equation}
and from our assumption (\ref{eq:rho_the_same_tilde}) this reduces
to 
\begin{equation}
\sum_{i,j}\sum_{\alpha,\beta}J_{i,j}^{\alpha,\beta}\rho_{i,j}^{\alpha,\beta}\left[\Psi_{0}\right]<\sum_{i,j}\sum_{\alpha,\beta}J_{i,j}^{\alpha,\beta}\rho_{i,j}^{\alpha,\beta}\left[\Psi_{0}\right]\label{eq:contrad_0}
\end{equation}
which is a contradiction. Therefore, the only possibility is that
the two expectation values are equal: 
\[
E_{0}\equiv\left\langle \Psi_{0}\left|\hat{H}\right|\Psi_{0}\right\rangle =\left\langle \tilde{\Psi}\left|\hat{H}\right|\tilde{\Psi}\right\rangle .
\]
However in that case $\left\langle \tilde{\Psi}\left|\hat{H}\right|\tilde{\Psi}\right\rangle $
is the absolute minimum $E_{0}$ and therefore $|\tilde{\Psi}\rangle$
is a ground state of $\hat{H}$, which contradicts our starting assumption.
Thus we conclude that the only state that reproduces the ground-state
correlator of $\hat{H}$ is the actual ground state of $\hat{H}$
(or one of its ground states, if the ground state of $\hat{H}$ happens
to be degenerate), \emph{quod erat demonstrandum}.$\square$ 

We note that our proof of Theorem~2 does not rely on having proved
Theorem~1. Theorem~2 is a simple consequence of the fact that the
expectation value of any Hamiltonian of the form (\ref{eq:H}) is
a sum of two-point correlators. Thus, if two states $|\Psi_{0}\rangle,|\tilde{\Psi}\rangle$
give the same correlators they must give the same expectation value.
Therefore, if $|\Psi_{0}\rangle$ minimizes the energy $|\tilde{\Psi}\rangle$
does too. We also stress that Theorem~2 is true even when we include
candidate wave functions such as $\Psi_{\text{f}}$ in Fig.~\ref{fig:Possible-relationships-between}
that are not derived from any Hamiltonian of the form (\ref{eq:H})
(which makes Theorem 2 stronger than Theorem 1). This means that unconstrained
searches in wave function space are guaranteed to be able to find
the true ground state.

Our last result offers the possibility to study systems for which
experimental magnetic neutron scattering data are available by working
directly with the wave function, without the need for a model Hamiltonian
(Fig.~\ref{fig:-Two-versions}~b). The same efficient encodings
of wave functions that have been developed to obtain the ground state
of a given model Hamiltonian could be used to find the wave function
that matches the experimental data. For instance, one could encode
the wave function in a neural network \citep{ANN1}, trained once
to reproduce the experimental data (instead of minimizing the energy
as done in Ref.\,\citealp{ANN1}). Alternatively, a quantum circuit
could be optimized to place the qubits in a quantum processor in a
state that reproduces the measurements. In this respect, we note that
the simulation of inelastic neutron scattering functions of single-molecule
magnets using a quantum processor (for known Hamiltonian) has already
been successfully demonstrated~\citep{Chiesa2019}. The approach
we propose would dispense with the model Hamiltonian and instead optimise
the scattering function directly. It would be similar to an evolutionary
variational eigensolver \citep{Rattew2020} except that, again, we
would not be minimizing the energy of a model Hamiltonian but would
be instead optimising the wave function to describe the experimental
data. Both neural networks (or, more generally, tensor networks \citep{Orus2019})
and quantum circuits can, in principle, generate any wave function.
Our theorem implies that any general-purpose optimization algorithm
will converge towards the right ground state (or another ground state
of the same model with the same correlators). Specifically, it guarantees
that convergence towards an unphysical wave function that reproduces
the data is not possible as there are no wave functions that can describe
the data and are not valid solutions to the problem at hand. This
is akin to the guarantee offered by the Rayleigh-Ritz variational
principle that no wave function can give an energy lower than the
true ground-state wave function. Once optimized, our neural network
or quantum circuit contain all the obtainable information about the
system's ground state and can straight-forwardly be used to predict
any other ground-state property.

To conclude we note some limitations of Theorem 2. Firstly, it relies
on the assumption that the physical system under investigation is
described by a Hamiltonian of the form in Eq.~(\ref{eq:H}). Systems
with itinerant electrons or with interaction terms involving three
or more spins at a time are therefore excluded. The generalization
of our results to such systems is left for subsequent work~\footnote{We note however that some notable Heisenberg-type models with higher-order
interaction terms emerge as perturbative approximations to models
that only feature 2-spin interactions e.g. the cyclic-exchange model
of quantum spin ice \citep{gingras2014quantum}. Such systems \emph{are
}covered by our theorems.}. Secondly, Theorem 2 establishes the existence of a fitness peak
at $\Psi_{0}$ but says nothing about its steepness. The peak could
be almost a plateau in some cases, which would complicate practical
applications. Investigating this for different models provides another
focus for future research. Finally, our theorems refer only to the
ground state (apart from the generalisation of Theorem 1 to excited
states in Appendix \ref{sec:Can-two-different}). Further generalizations
to states of thermodynamic equilibrium and to excited states are left
for future work. \emph{Note added.-- }the generalisation of Theorem\,1
to finite temperatures has been discussed recently by Murta and Fern\'andez-Rossier
\citep{Murta2022}. 
\begin{acknowledgments}
The author wishes to thank Silvia Ramos for asking the question that
motivated this work; Vivaldo Campo and Klaus Capelle for a stimulating
discussion of DFT for Heisenberg models; Gunnar M{\"o}ller and Tymoteusz
Tula for further useful discussions that informed the preparation
of this manuscript; and all of the above for further useful comments
on a draft. 
\end{acknowledgments}

\appendix

\section{Equivalence between correlators and the spin structure factor\label{sec:Equivalence-between-correlators}}

Here we show that, for systems with translational symmetry, the correlators
$\rho_{ij}^{\alpha,\beta}$ are unique functionals of the diffuse
magnetic neutron scattering function, or static spin structure factor,
$S_{\alpha,\beta}$$\left(\mathbf{q}\right)$, which can be determined
experimentally \citep{Lovesey1987,Zaliznyak2004} and is given by 

\begin{equation}
S_{\alpha,\beta}\left(\mathbf{q}\right)\equiv\frac{1}{N\hbar}\sum_{i,j}e^{i\mathbf{q}\cdot\left(\mathbf{R}_{i}-\mathbf{R}_{j}\right)}\left\langle \hat{S}_{i}^{\alpha}\hat{S}_{j}^{\beta}\right\rangle .\label{eq:S}
\end{equation}
This is not quite the same as a Fourier transform, in which case we
could say straightaway there is a one-to-one correspondence between
$S_{\alpha,\beta}\left(\mathbf{q}\right)$ and $\left\langle \hat{S}_{i}^{\alpha}\hat{S}_{j}^{\beta}\right\rangle ,$
but almost.  Again, let us proceed by \emph{reductio ad absurdum}.
First, we assume that there are two different correlation functions
that give the same scattering function. Let us designate these two
correlation functions as $\rho_{i,j}^{\alpha,\beta}$ and $\tilde{\rho}_{i,j}^{\alpha,\beta}$,
respectively. Our assumption is that the difference $\Delta_{i,j}^{\alpha,\beta}\equiv\rho{}_{i,j}^{\alpha,\beta}-\tilde{\rho}{}_{i,j}^{\alpha,\beta}\neq0.$
Since they give the same scattering function we have 
\begin{align*}
S_{\alpha,\beta}\left(\mathbf{q}\right) & =\frac{1}{N\hbar}\sum_{i,j}e^{i\mathbf{q}\cdot\left(\mathbf{R}_{i}-\mathbf{R}_{j}\right)}\rho_{i,j}^{\alpha,\beta}\\
 & =\frac{1}{N\hbar}\sum_{i,j}e^{i\mathbf{q}\cdot\left(\mathbf{R}_{i}-\mathbf{R}_{j}\right)}\tilde{\rho}_{i,j}^{\alpha,\beta}
\end{align*}
for all $\mathbf{q},\alpha,\beta.$ The last equality implies that
\begin{equation}
\sum_{i,j}e^{i\mathbf{q}\cdot\left(\mathbf{R}_{i}-\mathbf{R}_{j}\right)}\Delta_{i,j}^{\alpha,\beta}=0\text{ for all }\mathbf{q},\alpha,\beta.\label{eq:Delta_is_zero}
\end{equation}

Suppose that all magnetic sites are equivalent. Then the function
$\Delta_{i,j}^{\alpha,\beta}=\Delta^{\alpha,\beta}\left(\mathbf{R}_{i}-\mathbf{R}_{j}\right)$
and (\ref{eq:Delta_is_zero}) becomes 
\[
\sum_{\mathbf{R}}e^{i\mathbf{q}\cdot\mathbf{R}}\Delta^{\alpha,\beta}\left(\mathbf{R}\right)=0\text{ for all }\mathbf{q},\alpha,\beta
\]
which evidently implies $\Delta^{\alpha,\beta}\left(\mathbf{R}\right)=0$
for all $\mathbf{R}$ as the Fourier transform of a null function
is a null function which contradicts our original assumption, concluding
our argument. 

Suppose now that the magnetic sites are not equivalent. Nevertheless,
as long as we are dealing with a state with translational symmetry,
the function $\rho_{i,j}^{\alpha,\beta}$ will have to be periodic.
This periodicity can be established experimentally (for instance,
by magnetic neutron crystallography) and it is also straight-forward
to impose it on the wave function therefore we can restrict ourselves
to the assumption that $\tilde{\rho}_{i,j}^{\alpha,\beta}$ (and therefore,
also $\Delta_{i,j}^{\alpha,\beta}$ ) has the same periodicity \footnote{This is the basis for expressing the neutron scattering cross-section
in terms of the static structure factor $S\left(\mathbf{q}\right)$
in the case of crystals with lowered periodicity (e.g. through an
anti-ferromagnetic transitions), see \citep{Zaliznyak2004} {[}see
in particular the discussion in Section 3.5{]}.}. In practice this means that we can write the LHS of Eq.~(\ref{eq:Delta_is_zero})
in the following form: 
\begin{align*}
\sum_{i,j}e^{i\mathbf{q}\cdot\left(\mathbf{R}_{i}-\mathbf{R}_{j}\right)}\Delta_{i,j}^{\alpha,\beta}= & \mathcal{N}\sum_{i=1}^{M\times\mathcal{N}}\sum_{j=1}^{M}e^{i\mathbf{q}\cdot\left(\mathbf{R}_{i}-\mathbf{R}_{j}\right)}\Delta_{j}^{\alpha,\beta}\left(\mathbf{R}_{i}\right)\\
= & \mathcal{N}\sum_{j=1}^{M}e^{-i\mathbf{q}\cdot\mathbf{R}_{j}}f_{j}\left(\mathbf{q}\right)
\end{align*}
with 
\[
f_{j}\left(\mathbf{q}\right)=\sum_{i=1}^{M\times\mathcal{N}}e^{i\mathbf{q}\cdot\mathbf{R}_{i}}\Delta_{j}^{\alpha,\beta}\left(\mathbf{R}_{i}\right).
\]
Here $\mathcal{N}$ is the number of magnetic unit cells (repeating
units) and $M$ is the number of sites within a unit cell. Thus the
sum over $j$ runs over all the sites in the first unit cell while
the sum over $i$ runs over all the sites in the lattice. For the
expression $\sum_{j=1}^{M}e^{-i\mathbf{q}\cdot\mathbf{R}_{j}}f_{j}\left(\mathbf{q}\right)$
to vanish for all $\mathbf{q}$ we must have each of the $f_{j}\left(\mathbf{q}\right)$
for $j=1,2,\ldots,M$ vanish independently. But $f_{j}\left(\mathbf{q}\right)$
is the Fourier transform of $\Delta_{j}^{\alpha,\beta}\left(\mathbf{R}_{i}\right)$
therefore $\Delta_{j}^{\alpha,\beta}\left(\mathbf{R}_{i}\right)$
must vanish too for each $j=1,2,\ldots,M$. This means $\Delta_{i,j}^{\alpha,\beta}$
is identically zero, contradicting again our starting assumption. 

There is a third possibility, namely the system may not be periodic.
This applies, for example, when there is quenched disorder. In that
case the scattering function $S_{\alpha,\beta}\left(\mathbf{q}\right)$
is averaged over the disorder and is therefore insufficient to determine
the real-space correlator. The extent to which $S_{\alpha,\beta}\left(\mathbf{q}\right)$
constrains the system's ground state in that case should be an interesting
subject for future investigations. 

\section{\label{sec:Can-two-different}Extension of Theorem 1 to Excited States}

Here we extend Theorem 1 to excited states. Consider two Hamiltonians
$\hat{H},\hat{H}'$ of the Heisenberg type {[}Eq.~(\ref{eq:H}){]}
but with different sets of coupling constants given by $J_{i,j}^{\alpha,\beta}$
and ${J'}_{i,j}^{\alpha,\beta}$, respectively. Let us assume that
the ground-state correlator $\rho_{i,j}^{\alpha,\beta}\left[\Psi_{0}\right]$
is the same. In that case $|\Psi_{0}\rangle$ is a ground state of
both Hamiltonians due to Theorem 1. Let $|\Psi_{1}\rangle$ and $|{\Psi'}_{1}\rangle$
the first excited states of $\hat{H}$ and $\hat{H}'$, respectively.
We wish to prove that if these two states are different the corresponding
correlators are also different, $\rho_{i,j}^{\alpha,\beta}\left[\Psi_{1}\right]\neq\rho_{i,j}^{\alpha,\beta}\left[{\Psi'}_{1}\right].$
As with all the other proofs in this paper, we proceed by \emph{reductio
ad absurdum}. Let us assume that the opposite is true, in other words
$\rho_{i,j}^{\alpha,\beta}\left[\Psi_{1}\right]=\rho_{i,j}^{\alpha,\beta}\left[{\Psi'}_{1}\right].$
Then 
\begin{align}
E_{1}=\left\langle \Psi_{1}\left|\hat{H}\right|\Psi_{1}\right\rangle  & <\left\langle \Psi'_{1}\left|\hat{H}\right|\Psi'_{1}\right\rangle \label{eq:ineq1-2}\\
 & =\left\langle \Psi'_{1}\left|\hat{H}-\hat{H}'\right|\Psi'_{1}\right\rangle +\left\langle \Psi'_{1}\left|\hat{H}'\right|\Psi'_{1}\right\rangle \\
 & =\sum_{i,j}\sum_{\alpha,\beta}\left(J_{i,j}^{\alpha,\beta}-{J'}{}_{i,j}^{\alpha,\beta}\right)\rho_{i,j}^{\alpha,\beta}\left[\Psi_{1}'\right]+E_{1}'
\end{align}
where in writing the inequality we have made use of our assumption
that $|\Psi_{1}\rangle\neq|{\Psi'}_{1}\rangle$. We have also used
that both $|\Psi_{1}\rangle$ and $|{\Psi'}_{1}\rangle$ are orthogonal
to the shared ground state $|\Psi_{0}\rangle.$ Similarly 
\begin{align}
{E'}_{1}=\left\langle \Psi_{1}'\left|\hat{H}'\right|\Psi_{1}'\right\rangle  & <\left\langle \Psi_{1}\left|\hat{H}'\right|\Psi_{1}\right\rangle \label{eq:ineq2-1}\\
 & =\left\langle \Psi_{1}\left|\hat{H}'-\hat{H}\right|\Psi_{1}\right\rangle +\left\langle \Psi_{1}\left|\hat{H}\right|\Psi_{1}\right\rangle \\
 & =\sum_{i,j}\sum_{\alpha,\beta}\left({J'}{}_{i,j}^{\alpha,\beta}-J_{i,j}^{\alpha,\beta}\right)\rho_{i,j}^{\alpha,\beta}\left[\Psi_{1}\right]+E_{1},
\end{align}
with the same assumptions made above. Adding the two inequalities
we obtain 
\[
E_{1}+{E'}_{1}<\sum_{i,j}\sum_{\alpha,\beta}\left(J_{i,j}^{\alpha,\beta}-{J'}{}_{i,j}^{\alpha,\beta}\right)\left(\rho_{i,j}^{\alpha,\beta}\left[\Psi_{1}'\right]-\rho_{i,j}^{\alpha,\beta}\left[\Psi_{1}\right]\right)+E_{1}'+E_{1}
\]
Our assumption that $\rho_{i,j}^{\alpha,\beta}\left[\Psi_{1}'\right]=\rho_{i,j}^{\alpha,\beta}\left[\Psi_{1}\right]$
then leads to 
\[
E_{1}+{E'}_{1}<E_{1}'+E_{1}
\]
which is absurd, \emph{quod erat demonstrandum}. The argument can
be trivially extended to successive excited states. We can also extend
it in the same way as Theorem 1 to cover the case where the excited
state is degenerate (in other words, to show that if $|\Psi_{1}\rangle$
and $|\Psi_{1}'\rangle$ are degenerate excited states of $\hat{H}$
then both of them must also be degenerate states of $\hat{H}'$).$\square$\bibliographystyle{apsrev4-1}
\bibliography{00_bibliography_paper_on_correlators}

\end{document}